\documentclass[a4paper,11pt]{article}

\usepackage{amsmath}
\usepackage{amssymb}
\usepackage{amsthm}
\usepackage{array}
\usepackage{indentfirst}
\usepackage{graphicx}
\usepackage[utf8]{inputenc}
\usepackage[dvipsnames]{xcolor}
\usepackage{tabularx}
\usepackage{caption}
\usepackage{subcaption}
\usepackage{gensymb}
\usepackage[normalem]{ulem}
\usepackage{array}
\usepackage{enumitem}
\usepackage{soul}
  \soulregister{\cite}{7}
  \soulregister{\ref}{7}
  \soulregister{\eqref}{7}
  \soulregister{\textbf}{0}

\usepackage{amssymb}

\usepackage{jcappub}
\newcolumntype{Y}{>{\centering\arraybackslash}X}
\newcolumntype{W}[1]{>{\centering\arraybackslash\hsize=#1\hsize}X}

\definecolor{lightblue}{rgb}{1.,1.,1.}
\sethlcolor{lightblue}
\newcommand{\chg}[1]{\hl{#1}}

\hypersetup{colorlinks=true, linkcolor=blue, urlcolor=blue, citecolor=blue, linktocpage=true}

\begin{document}

\title{Ising noise filter: physics-informed filtering for particle detectors.}

\author[a, b]{I. Kharuk}

\affiliation[a]{Institute for Nuclear Research of the Russian Academy of Sciences,
\\ 60th October Anniversary Prospect 7a, Moscow, 117312, Russia}
\affiliation[b]{Moscow Institute of Physics and Technology,\\
Institutsky lane, 9, Dolgoprudny, Moscow region, 141700, Russia}

\emailAdd{ivan.kharuk@phystech.edu}

\abstract{
We present the Ising noise filter, a highly portable, graph-based filtering algorithm for early-stage background suppression in particle accelerators and astrophysical detectors. 
Standard noise rejection methods relying on track fitting suffer from severe combinatorial explosion. 
Our method bypasses this by mapping individual detector hits to a network of binary spins and minimizing an energy functional. \chg{The entire algorithm reduces to vectorized array operations, making it fast and easy to deploy.}
The interaction kernels are physics-informed, tailored to the underlying physics and geometry of the experiment. 
We demonstrate the efficacy of this approach in two distinct experimental regimes. 
Applied to the Baikal-GVD neutrino telescope the filter yields fast, standard-quality noise rejection with 97\% recall for astrophysical neutrinos. 
For the SPD detector at the NICA collider the filter attains \chg{96.7\% precision and 97.4\% recall} on a toy Monte Carlo sample. 
Furthermore, when combined with a Peterson--Hopfield network for track finding, our physics-informed coupling improves the TrackML score from 0.5 to 0.91.
}

\maketitle

\section{Introduction}

Suppression of background noise in detector modules is a ubiquitous challenge in modern particle physics experiments, encompassing both particle accelerators and astrophysical detectors. Depending on the detector design and the underlying physical environment, noise may originate from thermal fluctuations in photosensors, natural background luminescence, or electronic cross-talk. Because the noise fraction frequently dominates the physical signal, reliable and computationally efficient filtering is a prerequisite for data analysis.
 
The most common approach to noise rejection relies on track fitting~\cite{mankel2004pattern, strandlie2010track}: algorithms reconstruct particle trajectories and discard inconsistent detector module activations (hits). While powerful, this approach is computationally expensive due to the circular dependency between noise filtering and track finding. Alternative pre-filtering methods, such as simple geometric cuts, time-coincidence requirements, and machine-learning classifiers~\cite{Radovic2018, Albertsson2018}, offer various trade-offs in accuracy, speed, and interpretability.
 
In this paper, we introduce a different paradigm for early-stage noise suppression: an Ising-model-based~\cite{Ising1925} filter that evaluates the physical consistency of an event independent of global track reconstruction. By assigning a binary spin state (signal or noise) to every individual hit, we map the detector data onto a network of interacting spins. The coupling strengths between these spins are defined by physical interaction kernels encoding underlying physics and geometric specifics of the experiment. Through the minimization of the system's energy functional, correlated signal hits collectively support one another, while uncorrelated noise hits are isolated and suppressed. 

The Ising noise filter resolves the combinatorial explosion inherent in early-stage track finding. \chg{Its computational complexity scales as $O(N^2)$ in the number of hits, or $O(N k)$ with a
spatial cutoff, and converges in a few iterations. The algorithm contains no
sequential loops, which allows using it in a fully vectorized form, independent from $N$. This makes our approach a good fit for online or early-stage data processing.}

The energy-minimisation approach was applied previously in particle physics. The Peterson–Hopfield network~\cite{Peterson1989, Hopfield1982} applies a closely related Hamiltonian formulation to track finding in drift chambers, operating on oriented hit doublets rather than individual hits. Image denoising~\cite{Geman1984} is another example of this method. Our contribution is to show that by tailoring the interaction kernels to the specific physics of each experiment, this framework becomes a powerful and highly portable noise filter, which can be adopted to any physical experiment.
 
We demonstrate the efficacy of this method in two distinct experimental contexts. First, we apply it to Baikal-GVD, a cubic-kilometer neutrino telescope. By encoding Cherenkov light propagation into the hit couplings, the Ising filter achieves reconstruction quality comparable to standard algorithms while operating significantly faster. Second, we apply the method to the SPD detector at the NICA collider, where electronic noise and beam-gas interactions contribute roughly 55\% of all hits. The Ising filter significantly lowers the noise rate, thereby increasing the speed and accuracy of subsequent track fitting via Kalman filter~\cite{Fruhwirth1987} or Peterson–Hopfield network~\cite{Peterson1989}. Furthermore, we show that by adjusting track segment couplings to SPD-specific geometries, overall track reconstruction performance can be explicitly boosted. 
 
The paper is organised as follows. Section~\ref{baikal} introduces the Ising filter on the example of Baikal-GVD experiment. Section~\ref{spd} covers its application to SPD and the improvement of the Peterson track finder. Section~\ref{conclusion} summarizes the results and discusses broader applicability.

\section{Neutrino telescopes: Baikal-GVD}
\label{baikal}

\subsection{Detector and noise environment}

\begin{figure}[htbp]
 \centering
 \begin{subfigure}[b]{0.62\textwidth}
     \centering
     \includegraphics[width=\textwidth]{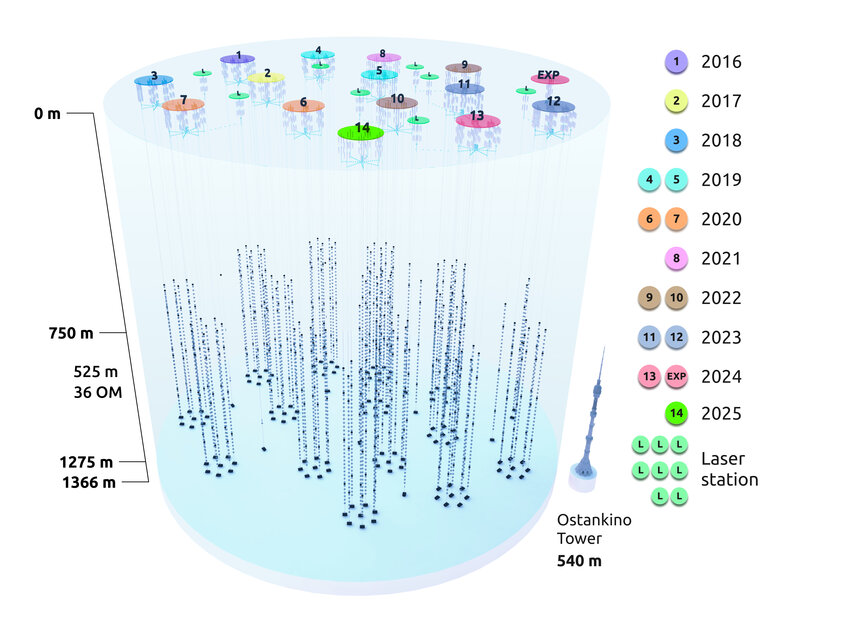}
     \caption{Baikal-GVD clusters layout.}
 \end{subfigure}
 \hfill
 \begin{subfigure}[b]{0.35\textwidth}
     \centering
     \includegraphics[width=\textwidth]{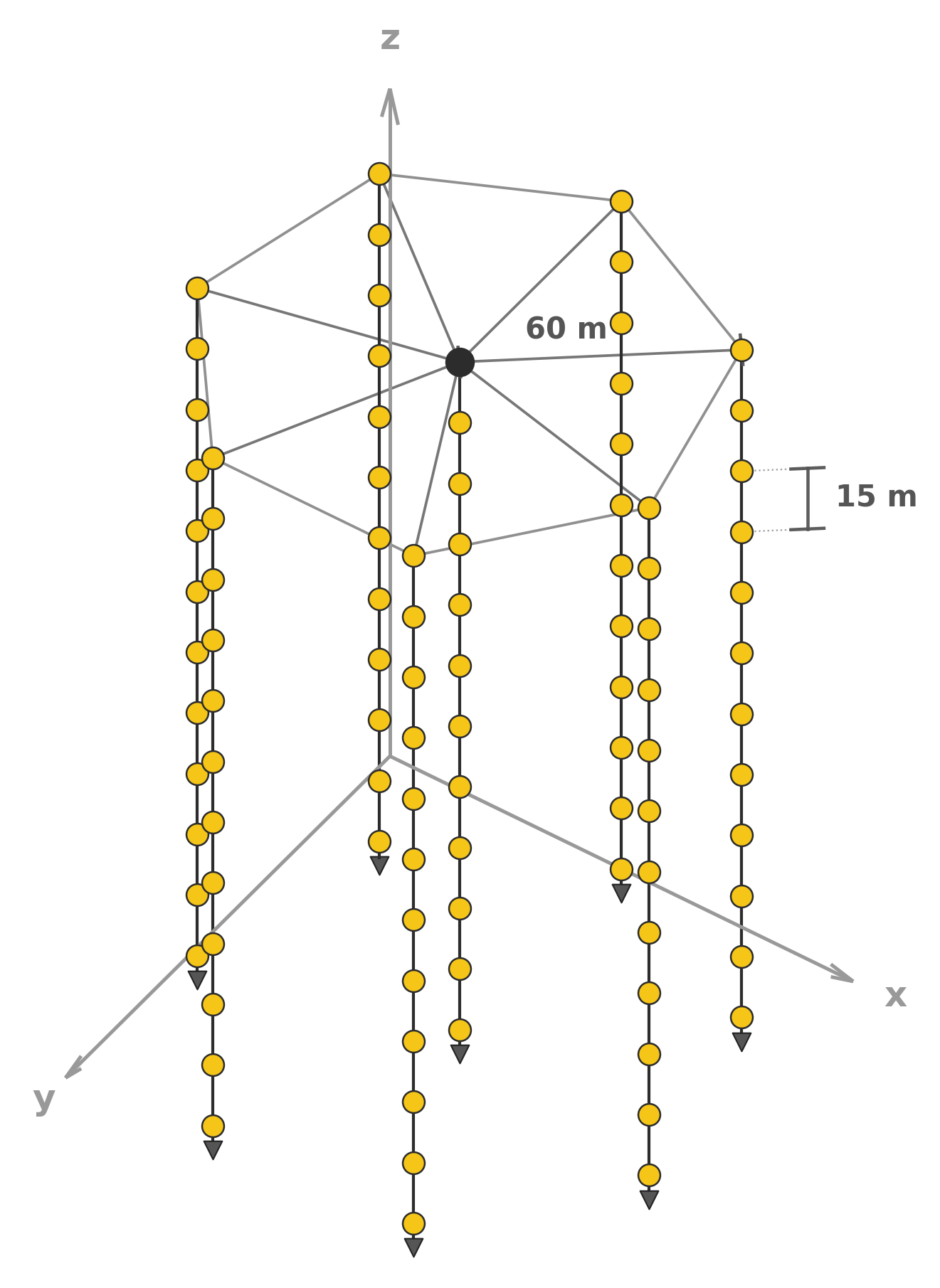}
     \caption{Single Baikal-GVD cluster.}
 \end{subfigure}
 \caption{Baikal-GVD design.}
 \label{baikal_detector}
\end{figure}

Baikal-GVD (Gigaton Volume Detector) is a cubic-kilometer neutrino telescope currently under construction in Lake Baikal, Russia~\cite{BaikalGVD2023status, BaikalGVD2023PRD}. Situated at a depth of 1366 m, its primary scientific program targets high-energy astrophysical neutrinos in the TeV–PeV range, encompassing point-source searches, diffuse flux measurements, and multi-messenger follow-ups of transient events. The detector's modular architecture, shown in figure \ref{baikal_detector}, is built from functional units called clusters. Each cluster has eight vertical strings, which are organized in an approximate regular heptagon with additional string at its center. Each string carries 36 optical modules (OMs) spaced 15 m apart vertically, with strings separated by roughly 60 m within a given cluster. These OMs are sensitive to single photons, recording both the arrival time and the integrated charge of each detected pulse. \chg{In this work we consider single-cluster events, in which all signal hits are contained within one cluster.}
 
\begin{figure}[htbp]
    \centering
    \includegraphics[width=0.75\textwidth]{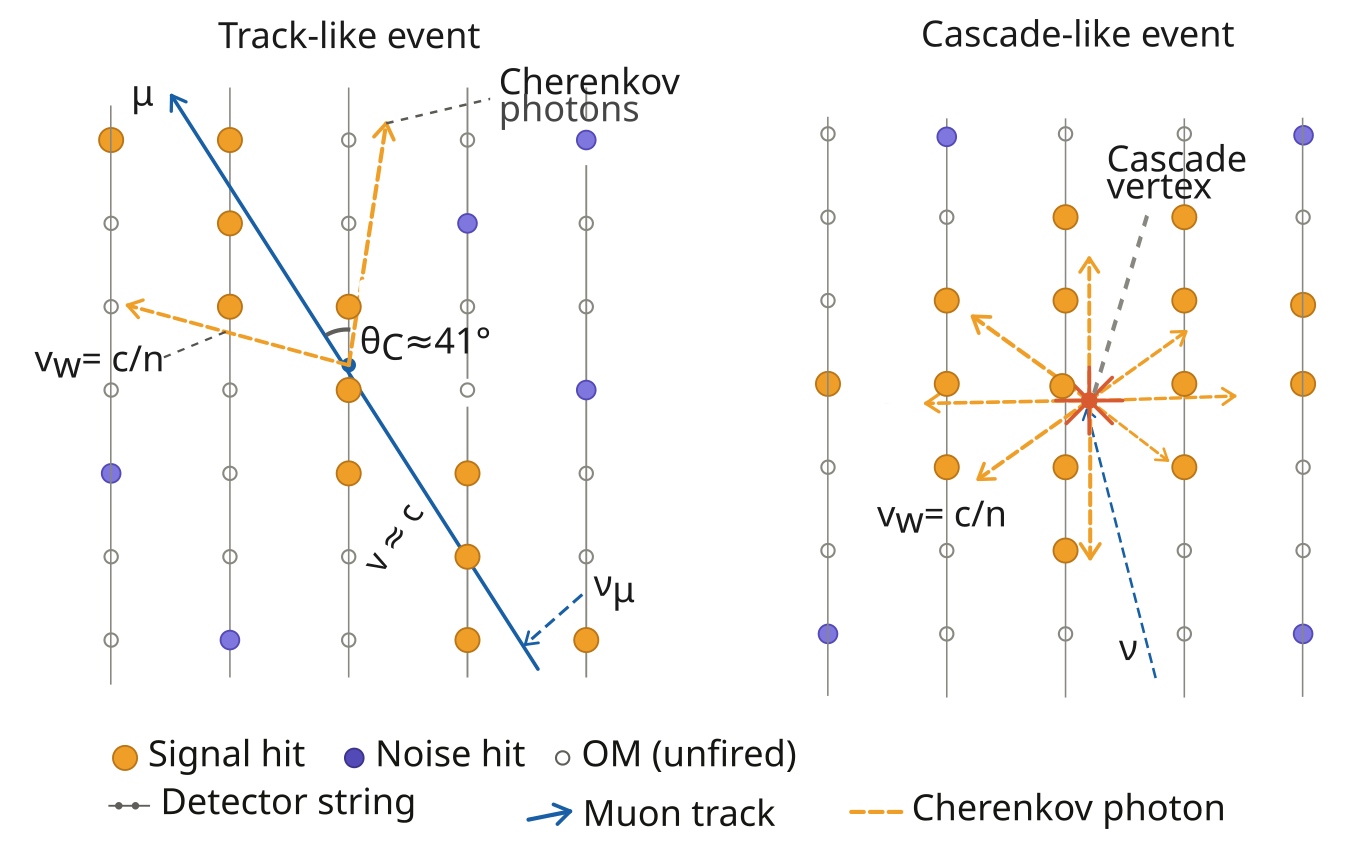}
    \caption{Illustration of neutrino-induced event topologies.}
    \label{event_topologies}
\end{figure}

Neutrino interactions within the instrumented water volume produce either hadronic and electromagnetic cascades or, in the case of charged-current muon-neutrino events, relativistic muon tracks. Both event topologies, illustrated in figure \ref{event_topologies}, emit Cherenkov radiation, with photons propagating at a speed of $v_w \approx c/1.35$ and at a characteristic Cherenkov angle of $\theta_C \approx 41^\circ$ relative to the particle's direction. Consequently, the OM signals originating from a single neutrino event are strongly correlated in space and time, strictly constrained by Cherenkov propagation dynamics.
 
The dominant source of background in Baikal-GVD is natural water luminescence~\cite{BaikalLuminescence2019}, which generates uncorrelated single-photon hits distributed across the detector. Because these noise hits can constitute up to 90\% of the collected data, an effective filtering is required for accurate event reconstruction. The standard approach to noise rejection for track-like events relies on a scan-fit algorithm~\cite{BaikalGVD2021hitfind,BaikalGVD2021muon}. This method scans over possible muon incoming directions using a finite angular step, searching for maximal cliques of hits that satisfy directed causality conditions. The directions yielding the largest cliques are then used as seeds for downstream reconstruction. While reliable, this approach becomes computationally prohibitive, particularly for events containing more than 150 hits.
 
\subsection{Mapping event to Ising model}
 
\chg{The Ising model~\cite{Ising1925} is a paradigmatic model of statistical physics. It describes a collection of binary spins $s_i \in \{+1, -1\}$ residing on the nodes of a graph and interacting through pairwise couplings $J_{ij}$, with an energy $E = -\sum_{i<j} J_{ij}\, s_i s_j - \sum_i h_i\, s_i$, where the local fields $h_i$ bias individual spins. For ferromagnetic couplings ($J_{ij} \geq 0$) the energy is lowered when connected spins share the same value, so low-energy configurations organize into extended domains of aligned spins.}

We frame event filtering as a graph-based binary classification problem. Let an event be a fully connected graph where each recorded hit is a node, defined by its position $\mathbf{r}_i$, time $t_i$, and charge $q_i$. The goal is to assign a class label $s_i \in \{+1, -1\}$ to each node, where $+1$ indicates signal and $-1$ indicates noise. \chg{Only channels that register a pulse form nodes, a channel with no recorded signal doesn't enter the graph.} The edge weights $J_{ij} \geq 0$ represent the interaction strength between nodes, quantifying the physical consistency of two hits originating from the same physical process. \chg{We aim at defining the dynamics of the system so that compact ``ferromagnetic'' domain of aligned $s = +1$ spins corresponds to the correlated hits of a genuine physical event, while uncorrelated noise hits are driven to $s = -1$.}
 
To capture the complex event topologies in Baikal-GVD, $J_{ij}$ is constructed as a linear combination of three physically motivated interaction kernels:
 
\textbf{Space-time proximity.} The broadest prior is that hits from the same neutrino event should be close in both space and time:
\begin{equation}
  J^{\mathrm{st}}_{ij}
  = \exp\!\left(-\frac{|\mathbf{r}_i - \mathbf{r}_j|}{\lambda_s}\right)
    \exp\!\left(-\frac{|t_i - t_j|}{\lambda_t}\right).
  \label{eq:Jst}
\end{equation}
Here $\lambda_s$ and $\lambda_t$ are characteristic length and time scales,
respectively.  This kernel assigns high weight to any correlated pair without
committing to a particular event topology.
 
\textbf{Track consistency.} Relativistic muons traverse the detector at speeds approaching $c$. Pairs of hits laying on a line approximately parallel to the muon track satisfy $|t_i - t_j| \approx |\mathbf{r}_i - \mathbf{r}_j|/c$.  We encode this correlation as:
\begin{equation}
  J^{\mathrm{tr}}_{ij}
  = \exp\!\left(
      -\frac{\bigl| |t_i - t_j| - |\mathbf{r}_i - \mathbf{r}_j|/c \bigr|}
            {\lambda_{\mathrm{tr}}}
    \right).
  \label{eq:Jvac}
\end{equation}

\textbf{Cherenkov-cone consistency.} For cascade events, and more generally for photons propagating through the medium, the relevant speed is $v_w = c/n$.  The corresponding causality condition yields:
\begin{equation}
  J^{\mathrm{wat}}_{ij}
  = \exp\!\left(
      -\frac{\bigl| |t_i - t_j| - |\mathbf{r}_i - \mathbf{r}_j|/v_w \bigr|}
            {\lambda_{\mathrm{wat}}}
    \right).
  \label{eq:Jwat}
\end{equation}
This rewards pairs consistent with photon propagation from a localized emission point.
 
The total coupling is a weighted sum:
\begin{equation}
  J_{ij} = c_{\mathrm{st}}\, J^{\mathrm{st}}_{ij}
          + c_{\mathrm{tr}}\, J^{\mathrm{tr}}_{ij}
          + c_{\mathrm{wat}}\, J^{\mathrm{wat}}_{ij} \;.
  \label{eq:Jtot}
\end{equation}
with coefficients $c_\cdot$ tuned on Monte Carlo simulations for optimal metrics.
 
In addition to the pairwise couplings, the measured charge $q_i$ provides hit-level information: a higher-amplitude hit is more likely to belong to the signal class. This is incorporated as an effective external local field, proportional to charge:
\begin{equation}
  h_i = q_{\mathrm{w}} \cdot \log\left(1 + \frac{q_i}{1~\text{p.e.}}\right) \;,
  \label{eq:hcharge}
\end{equation}
where \chg{$q_i/1~\text{p.e.}$ is the dimensionless pulse charge expressed in photoelectrons, and} $q_{\mathrm{w}}$ is a tunable parameter.
 
Finally, we introduce the external ``magnetic'' $\lambda$ field that pulls all spins to the $s=-1$ state. \chg{Within the analogy with the Ising model, it sets the boundary between the ordered signal domain and the disordered noise background, akin to a field-driven transition between ordered and disordered phases.}

Summing up, the full energy of the system reads:
\begin{equation}
\begin{split}
 E(\mathbf{s}) = &E_{\mathbf{align}} + E_{\mathbf{q}} + E_{\mathbf{ext}} \,, \\ ~~ E_{\mathbf{align}} = -\sum_{i < j} J_{ij}\, s_i s_j \,, ~~ &E_{\mathbf{q}} = - \sum_i h_i\, s_i \,, ~~ E_{\mathbf{ext}} = \sum_i \lambda s_i \;.
  \label{eq:energy}
\end{split}
\end{equation}
$E_{\mathbf{align}}$ is the ferromagnetic Ising Hamiltonian: since all couplings satisfy $J_{ij} \geq 0$, the energy is minimized when strongly coupled hits align to the same spin state. $E_{\mathbf{q}}$ shifts the energy balance in favor of spins alignment with high-amplitude hits, and $E_{\mathbf{ext}}$ pulls spins to the $s=-1$ state.
 
To separate signal from noise, all spins are initialized to $s=+1$. We apply a \chg{synchronous} spin update rule: a hit remains in the signal state if its local support ($\sum_{j} J_{ij} + h_i$) exceeds the global threshold $\lambda$. Otherwise, it flips to $s=-1$. This update is iterated 2–5 times over all hits until the system settles into a local energy minimum.
 
\begin{figure}[htbp]
  \centering
    \includegraphics[width=0.85\textwidth]{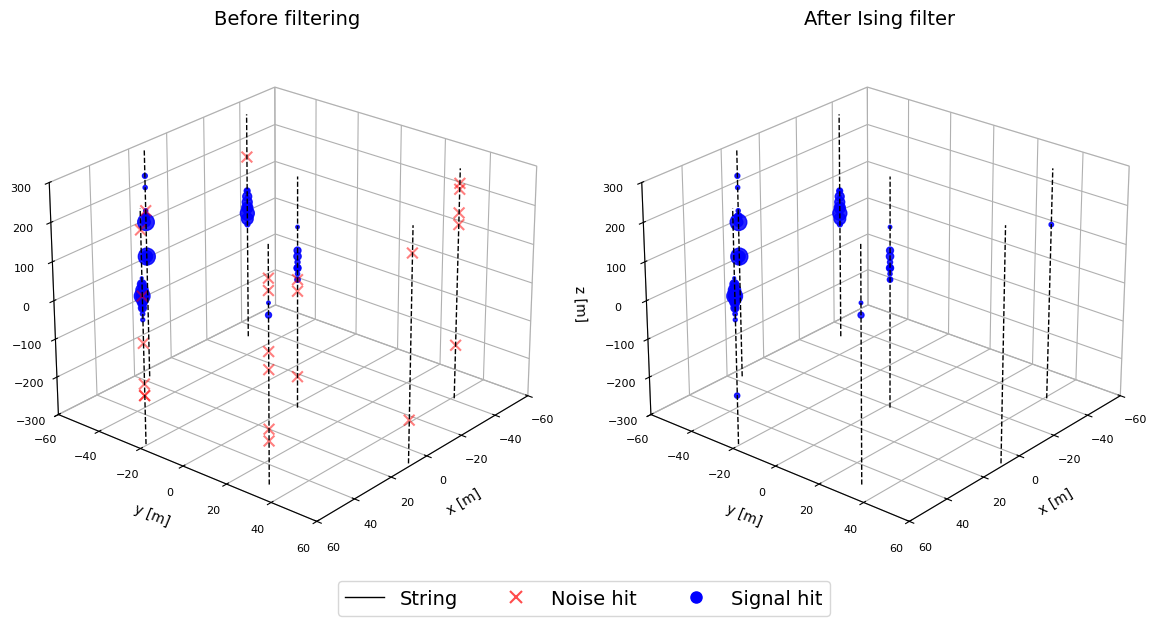}
    \caption{Illustration of applying Ising filter for noise filtering on Baikal-GVD data. Marker size for signal hits is proportional to the charge registered by OM.}
    \label{baikal_ising}
\end{figure}

The performance of the Ising filter was evaluated using muon neutrino and extensive air showers Monte Carlo (MC) datasets provided by the Baikal-GVD collaboration. We focused our optimization on the high-energy astrophysical neutrino sample, as these events are of primary physical interest. The filter hyperparameters were optimized to maximize the $F_1$ score,
\begin{equation}
    F_1 = \frac{2PR}{P+R} \;,
\end{equation}
ensuring a balanced trade-off between the retention of genuine signal hits (recall, $R$) and the rejection of background noise (precision, $P$). \chg{The optimized values used throughout this work are $\lambda_s = 35$~m, $\lambda_t = 50$~ns, $\lambda_{\mathrm{tr}} = 10$~ns, $\lambda_{\mathrm{wat}} = 20$~ns, $c_{\mathrm{st}} = 5$, $c_{\mathrm{tr}} = 2$, $c_{\mathrm{wat}} = 0.8$, $q_{\mathrm{w}} = 0.8$, and a global threshold $\lambda = 4$.} The resulting metrics are presented in table \ref{table_hitselection}, and a representative event display before and after the application of the Ising filter is shown in figure \ref{baikal_ising}. The Ising filter achieves 89\% event selection efficiency for the astrophysical neutrino sample with the standard Baikal-GVD event quality cuts: there are at least 8 identified signal hits on at least 2 strings. \chg{At this operating point the noise-rejection accuracy of the Ising filter matches that of the standard scan-fit algorithm~\cite{BaikalGVD2021hitfind, BaikalGVD2021muon}, while avoiding its combinatorial cost.}
 
\begin{table}
\begin{tabularx}{\textwidth}{|W{0.7}|W{1.1}|W{1.1}|W{1.1}|}
 \hline
~ & Extensive Air showers & Atmospheric neutrino &  Astrophysical neutrino \\
\hline
Precision & 90\% & 90\% & 91\%  \\
\hline
Recall & 93\% & 95\% & 97\%  \\
\hline
\end{tabularx}
\caption{Precision and recall metrics for events with at least 8 signal hits on at least 2 strings.}
\label{table_hitselection}
\end{table}
 
\chg{We have verified that greedy, single-spin and synchronous update rules yields identical precision and recall within statistical uncertainties. This proves that it is safe to use vectorized operations, replacing $N^2$ operations by a single matrix-vector multiplication, enabling efficient execution on modern hardware.}

\chg{In the canonical Ising model the spins are evolved at a finite temperature $T$: a flip that raises the energy by $\Delta E$ is still accepted with probability $\exp(-\Delta E / T)$~\cite{Glauber1963}, which allows the dynamics to climb out of shallow local minima. Simulated annealing~\cite{Kirkpatrick1983} exploits this by gradually lowering $T$ towards the global energy minimum. Our deterministic update rule corresponds to the zero-temperature limit $T \to 0$. This choice is deliberate: since the global minimum of Eq.~\eqref{eq:energy} is the trivial all-noise state $s_i = -1$, any finite temperature or annealing schedule would pull the system away from the physically meaningful, metastable signal configuration and degrade performance.}

\chg{We compared Ising filter performance against the neural-network-based approach of \cite{kharuk2023rejecting}, which was using U-Net-like architecture for signal hits identification. The neural network achieves 99\% precision and 96\% recall on Monte Carlo data, while Ising filter attains comparable recall (97\% for astrophysical neutrinos) with somewhat lower precision (91\%). However, the Ising filter offers substantial practical advantages. Each synchronous update of the Ising filter is analogous to a single layer of a neural network, yet the filter converges in 2–5 iterations and involves far fewer parameters. The total number of floating-point operations per event is therefore orders of magnitude smaller, making the Ising filter suitable for online or resource-constrained environments.}

\section{Accelerator physics: SPD at NICA}
\label{spd}
 
\subsection{Detector and noise environment}
 
The Spin Physics Detector (SPD) is the principal detector of the NICA collider~\cite{NICA2023}, currently under construction at JINR (Dubna, Russia)~\cite{SPD_TDR2024,SPD_CDR2021}. We will focus on SPD straw barrel tracker, which is aimed to register tracks of particles produced in the collider. Unlike Baikal-GVD case, there is no precise sub-nanosecond timing information available, hence background suppression must rely entirely on spatial and geometric constraints.
 
\begin{figure}[htbp]
  \centering
    \includegraphics[width=0.85\textwidth]{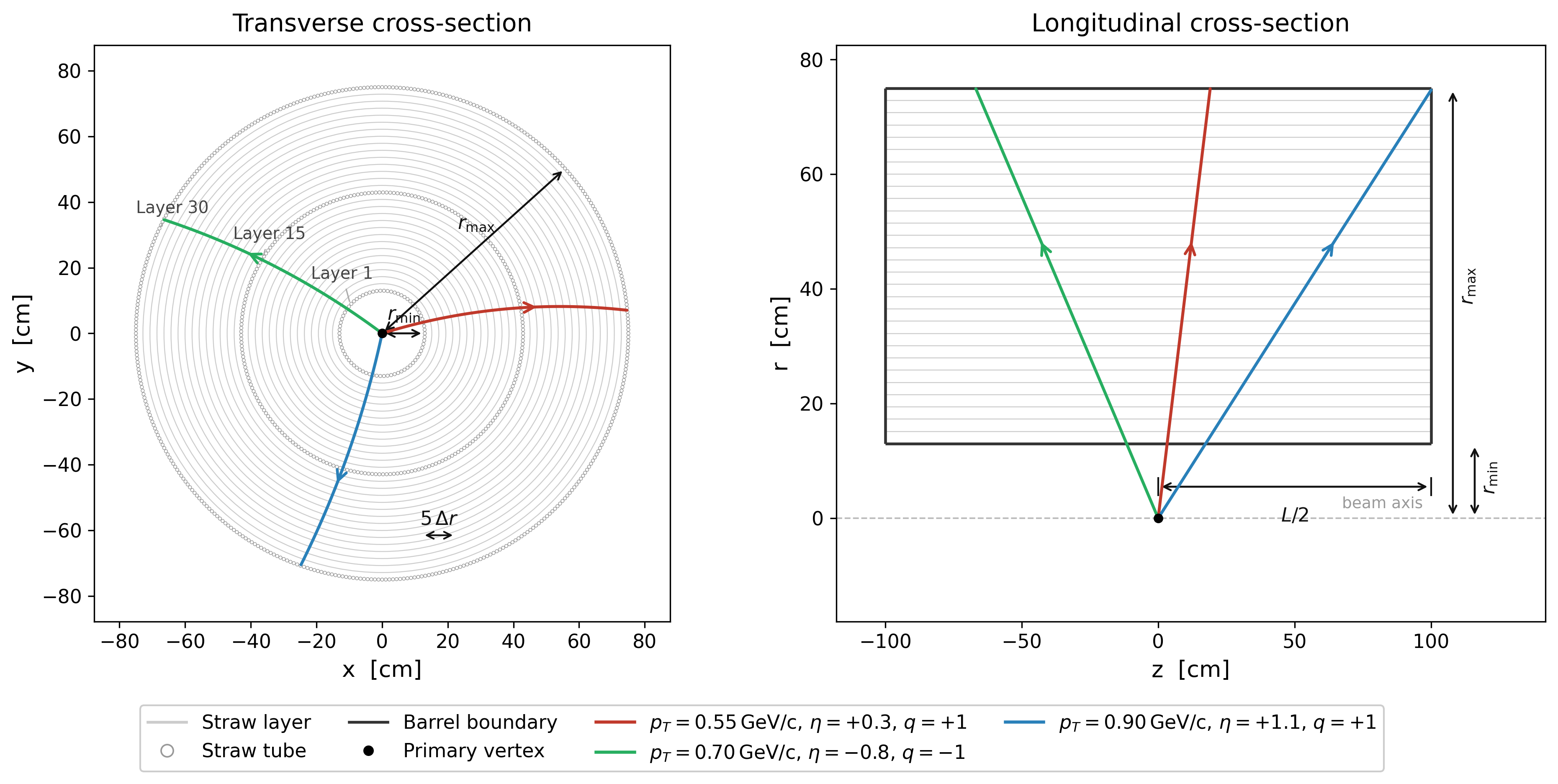}
    \caption{SPD detector design.}
    \label{spd_design}
\end{figure}

The SPD straw barrel tracker is designed to measure charged-particle tracks in a 1\,T solenoidal magnetic field. The barrel comprises 30 concentric cylindrical layers at radii between 13\,cm and 75\,cm, each constructed from 10\,mm straw tubes, figure~\ref{spd_design}. \chg{At the trigger stage considered here the drift time is not yet available, so the transverse position is known only at the straw-cell scale ($\sim 1\,$cm). The $z$-resolution from charge division ($\sim 3\,$cm) is already available at this stage, and both the filter and the Monte Carlo sample adopt these trigger-stage resolutions.} Charged particles traverse the tracker on helical trajectories whose transverse curvature is set by the particle's transverse momentum $p_T$ and the magnetic field strength. Because the $z$-resolution is poor relative to the layer spacing, reconstruction algorithms rely primarily on the more precisely measured transverse coordinates.
 
In SPD operating conditions, electronic noise, beam–gas interactions, and secondary particles produce a heavily contaminated data sample. At a channel occupancy of 2\% and an average of five primary tracks per event, noise accounts for approximately 55\% of all registered hits. Standard track identification algorithms, such as Peterson–Hopfield network~\cite{Peterson1989,Hopfield1982} and Kalman filter~\cite{Fruhwirth1987}, are combinatorially sensitive to hit multiplicity. A fast pre-filtering step that suppresses noise before reconstruction begins is therefore essential.

\subsection{Ising filtering for SPD}
\label{sec:spd-ising}

For the SPD detector, the Ising filter is built entirely on the geometry of helical tracks in a uniform magnetic field. The graph representation of events is similar to the Baikal-GVD case: each hit is a node characterised by its position $\mathbf{r}_i = (x_i, y_i, z_i)$ and the layer index $\ell_i \in \{1, \ldots, 30\}$. \chg{As in the Baikal-GVD case, only fired channels constitute nodes.} Edges are restricted to hit pairs within a cutoff radius $r_{\mathrm{cut}} = 11 \,$cm ($\approx$ 5 layer spacings), which encompasses the geometrically meaningful neighbourhood of any hit while eliminating computationally costly, physically uninformative long-range contributions. All pairs outside this radius are assigned $J_{ij} = 0$.
 
We define three physically motivated interaction kernels for identifying signal hits:
 
\textbf{Three-dimensional spatial proximity.} The simplest prior is that
consecutive hits from the same track on adjacent layers are close in 3-D space:
\begin{equation}
  J^{\mathrm{space}}_{ij}
  = \exp\!\left(-\frac{|\mathbf{r}_i - \mathbf{r}_j|}{\lambda_s}\right).
  \label{eq:Jspace}
\end{equation}
Signal hit pairs on adjacent layers are separated by 3--8\,cm, while
noise hits are randomly scattered across the detector.
 
\textbf{Transverse-plane proximity.} For tracks at large pseudorapidity
$|\eta| \gg 1$, the $z$-displacement between adjacent-layer hits is much
larger than the transverse displacement. In this regime the 3-D distance
is dominated by $\Delta z$, and $J^{\mathrm{space}}_{ij}$ underestimates
the correlation between genuine neighbours. A complementary kernel
measuring proximity in the transverse $(x,y)$ plane recovers sensitivity
to high-pseudorapidity tracks:
\begin{equation}
  J^{\mathrm{xy}}_{ij}
  = \exp\!\left(-\frac{\sqrt{(x_i-x_j)^2 + (y_i-y_j)^2}}{\lambda_{xy}}\right).
  \label{eq:Jxy}
\end{equation}
 
\textbf{Radial-chord alignment.} In the straw barrel geometry, consecutive hits from the same track on adjacent layers sit at nearly the same azimuthal angle $\phi$. Consequently, their transverse chord is nearly radial. We introduce a kernel that penalizes tangential chords, where $\cos\alpha_{ij} = |\hat{c}_{ij} \cdot \hat{r}_\mathrm{mid}|$ measures the alignment of the transverse chord $\hat{c}_{ij}$ with the radial vector at the chord's midpoint $\hat{r}_{\mathrm{mid}}$:
\begin{equation}
  J^{\mathrm{tang}}_{ij}
  = \exp\!\left(-\frac{1 - \cos\alpha_{ij}}{\lambda_{\mathrm{tang}}}\right)
  \label{eq:Jtang}
\end{equation}

The total SPD coupling is a weighted sum of these kernels:
\begin{equation}
  J_{ij} = c_{\mathrm{space}}\, J^{\mathrm{space}}_{ij}
          + c_{\mathrm{xy}}\,    J^{\mathrm{xy}}_{ij}
          + c_{\mathrm{tang}}\,  J^{\mathrm{tang}}_{ij} \;.
  \label{eq:JtotSPD}
\end{equation}

The filter was optimized and evaluated on a toy self-produced Monte Carlo sample simulating the SPD barrel tracker, comprising 500 events. Each event contains on average five primary tracks with momenta drawn uniformly from $p \in [0.5,\,5.0]$\,GeV/$c$ and polar angles spanning the full tracker acceptance. Helical propagation through all 30 layers is computed analytically in the nominal field of $B = 1$\,T, and hits are placed at the geometric intersection of each helix with the straw-layer surfaces. A channel occupancy of 2\% is applied as uncorrelated noise, resulting in approximately \chg{55\%} noise hits per event.
 
\begin{figure}[htbp]
    \centering
    \includegraphics[width=0.9\textwidth]{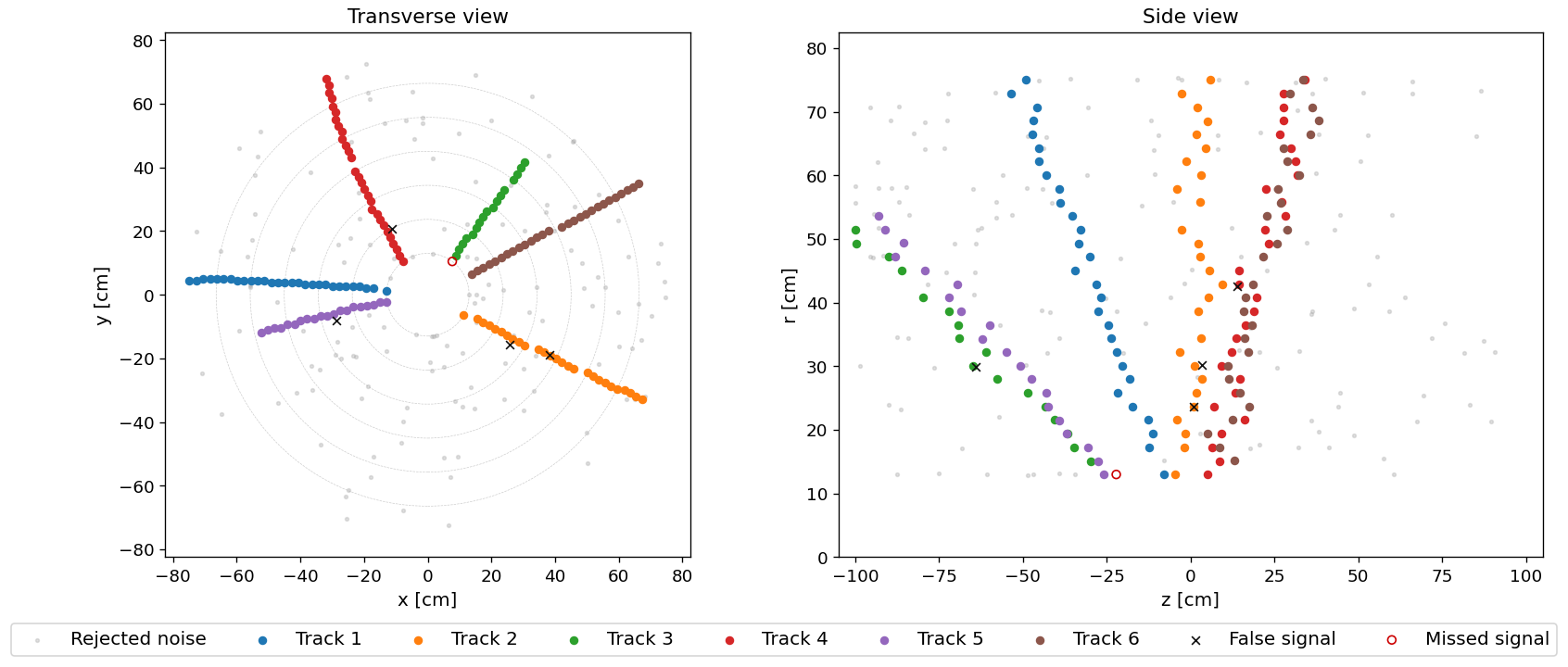}
    \caption{Illustration of Ising filter for the SPD detector. Tracks are denoted according to ground truth Monte Carlo data.}
    \label{spd_ising_example}
\end{figure}

With optimized hyperparameters, the filter achieved a precision of $96.7\%$ and a recall of $97.4\%$, yielding an overall $F_1$ score of $0.97$. An example of noise filtering results is shown in figure \ref{spd_ising_example}. In practice, it might be beneficial to prioritize recall over precision to keep all signal hits, and filter the remaining noise hits at later stages of data analysis.
 
\chg{In the actual SPD read-out, several collision events will be recorded within a single data frame. To assess the Ising filter performance under more realistic conditions, we estimate the metrics for a read-out containing 5 overlapping events, with the Ising filter hyperparameters optimized for this specific occupancy. In this regime the filter achieves a precision of $92.5\%$ and a recall of $96.3\%$. The degradation relative to the single-event case is modest, reflecting the robustness of the physics-informed kernels to increased hit multiplicity.}
 
\chg{These results can be compared with the Point Cloud Transformer (PCT) architecture developed for the same SPD signal segmentation task~\cite{nikolskaiia2024point}. The PCT employs self-attention modules with $O(N^2)$ algorithmic and memory complexity and requires a voxelization preprocessing step. While the PCT is a powerful general-purpose architecture, the Ising filter offers competitive precision and recall with significantly fewer parameters and lower computational cost: each synchronous iteration is analogous to a single neural-network layer, and convergence is reached within a few iterations. This makes the Ising filter particularly attractive for online filtering, where inference latency and memory footprint are critical constraints.}

\chg{We benchmarked the Ising filter as a pre-processing stage for a downstream extended Kalman filter (EKF) track fit using a common reference implementation on the same hardware, evaluated on realistic five-event pile-up read-out frames. The standalone Ising filter processes one frame in $\sim$60~ms. By removing background hits before track seeding it reduces the number of seeds and Kalman updates, lowering the total track-fitting time by a factor of 2. It also suppresses fake tracks roughly six-fold, from approximately 38 to 6 per frame, while leaving the tracking efficiency unchanged ($91\% \to 90\%$) and improving the average hit purity ($84\% \to 87\%$).}

\subsection{Track finding with the Peterson--Hopfield network}
 
Ising-filter noise suppression reduces the hit count sufficiently that computationally demanding track-finding algorithms become practical. In this subsection we demonstrate that the Peterson--Hopfield network~\cite{Peterson1989, Hopfield1982, StimpflAbele1991, Bures2024} --- a Hopfield-type neural network formulated on oriented hit pairs --- can be applied to the Ising-filtered SPD data. Furthermore, its performance is substantially improved by replacing the standard geometry-agnostic excitatory coupling with SPD-specific constraints derived directly from helical track geometry.
 
The Peterson--Hopfield network operates not on individual hits, but on doublets: oriented pairs of hits $(i \to j)$ on adjacent layers with $\ell_j = \ell_i + 1$. Each doublet $d$ is assigned a binary spin, $s_d = 1$ or $s_d = 0$. The dynamics of spins, defined below, is such that doublets with $s_d = 1$ are classified as active segment of a reconstructed track. The network finds a configuration in which active doublets form geometrically consistent, non-overlapping chains — one per track — from which individual tracks can be extracted as connected components of the doublet graph.
 
The network energy has two contributions. An \emph{excitatory} coupling $T_{d_1 d_2} > 0$ is assigned to consecutive doublet pairs $(i \to j)$ and $(j \to k)$ that share a common middle hit, rewarding geometrically consistent track segments. An \emph{inhibitory} coupling $W_{d_1 d_2} < 0$ penalises pairs of doublets that share a hit but are \emph{not} consecutive: since a single hit can belong to at most one track, two such doublets being simultaneously active raises the energy, suppressing ambiguous hit assignments. Together, these two terms drive the network towards a state where each active doublet belongs to exactly one smooth, unambiguous track. The total energy is:
\begin{equation}
  E_\mathrm{P} = -\frac{1}{2}\sum_{d_1 \neq d_2} T_{d_1 d_2}\, s_{d_1} s_{d_2}
               + \frac{1}{2}\sum_{d_1 \neq d_2} |W_{d_1 d_2}|\, s_{d_1} s_{d_2}
               + \mu \sum_d \left(s_d - N\right)^2 \;,
  \label{eq:PetersonEnergy}
\end{equation}
where $N$ is the total number of doublets. The last term enforces the network to assign track labels, $s_d=1$, to at least some of the doublets.
 
To find the ground state, doublets spins are updated several times till convergence. Individual tracks are subsequently extracted as connected components of the doublet graph, with optional post-processing to resolve track merging ambiguities~\cite{Peterson1989}. For better performance, one can apply simulated annealing~\cite{Kirkpatrick1983} and mean field theory approach~\cite{PetersonAnderson1987}. 
 
The standard Peterson formulation relies on a geometry-agnostic cosine-kink excitatory coupling to link consecutive doublets $(i \to j)$ and $(j \to k)$:
\begin{equation}
  T^{\mathrm{std}}_{d_1 d_2}
  = \max\!\left(0,\; \cos\theta^{ijk}_\mathrm{kink}\right), \qquad
  \cos\theta^{ijk}_\mathrm{kink}
  = \frac{(\mathbf{r}_j - \mathbf{r}_i)\cdot(\mathbf{r}_k - \mathbf{r}_j)}
         {|\mathbf{r}_j - \mathbf{r}_i|\,|\mathbf{r}_k - \mathbf{r}_j|} \;.
  \label{eq:Tstd}
\end{equation}
Triplets from the same track on a helix have a small kink angle ($\theta_\mathrm{kink} \approx 0$), giving $T^{\mathrm{std}} \approx 1$, while spurious triplets with random orientations score lower.
 
We propose an improved excitatory coupling that encodes two constraints specific to helical track geometry in a solenoidal field.
 
\textbf{Transverse bending angle.} On a helix in a uniform field, the transverse direction of a track changes smoothly and uniformly between adjacent layer pairs.  The change in transverse azimuthal angle between consecutive chords $(i \to j)$ and $(j \to k)$ is:
\begin{equation}
  \Delta\phi^{ijk}
  = \arg\!\left(\exp\!\left[i\left(\phi_{jk} - \phi_{ij}\right)\right]\right),
  \qquad
  \phi_{ab} = \mathrm{arctan}\left((y_b - y_a) / (x_b - x_a)\right) \;.
  \label{eq:deltaphi}
\end{equation}
Monte Carlo measurements on Ising-filtered hits give $\langle|\Delta\phi|\rangle = 0.17$\,rad for same-track triplets and $\langle|\Delta\phi|\rangle = 0.86$\,rad for fake triplets.
 
\textbf{$z$-slope consistency.} For a helix with polar angle $\theta$, corresponding to pseudorapidity $\eta = -\ln\tan(\theta/2)$, the $z$-displacement per unit layer index is $|dz/d\ell| = R_\mathrm{layer}\sinh\eta$, a conserved quantity along the track.
Consequently, a helix projects onto a straight line in the $z$-$r$ plane, and the signed $z$-slopes of the two constituent doublets of a triplet must agree. We therefore penalise the change in $z$-slope between the two half-segments $(i \to j)$ and $(j \to k)$:
\begin{equation}
  \delta\!\left(\frac{dz}{d\ell}\right)_{ijk}
  =
  \left|
    \frac{z_k - z_j}{\ell_k - \ell_j}
    -
    \frac{z_j - z_i}{\ell_j - \ell_i}
  \right| \;.
  \label{eq:zslope}
\end{equation}
This quantity vanishes exactly for a perfect helix regardless of $|\eta|$, so high-pseudorapidity tracks — which have a large but constant $z$-slope — are accepted on equal footing with central tracks. Only triplets assembled from geometrically inconsistent hit pairs, such as random noise combinations, produce large $\delta(dz/d\ell)$.
 
Both constraints are combined into total excitatory coupling:
\begin{equation}
  T^{\mathrm{SPD}}_{d_1 d_2}
  = \exp\!\left(
      -\frac{|\Delta\phi^{ijk}|}{\lambda_{\mathrm{bend}}}
      - \frac{1}{\lambda_z}\,\delta\!\left(\frac{dz}{d\ell}\right)_{ijk}
    \right) \;,
  \label{eq:Tspd}
\end{equation}
where $\lambda_{\mathrm{bend}}$ and $\lambda_z$ are scale parameters tuned by hyperparameter optimization. We do not use the cosine-kink excitatory coupling since transverse bending angle kernel better captures SPD track specifics.
 
The inhibitory coupling is unchanged from the standard formulation: doublets that share a hit but are not consecutive are penalised by a fixed negative weight $W_{d_1 d_2} = -\lambda_{\mathrm{inh}}$.
 
\begin{figure}[htbp]
  \centering
    \includegraphics[width=0.8\textwidth]{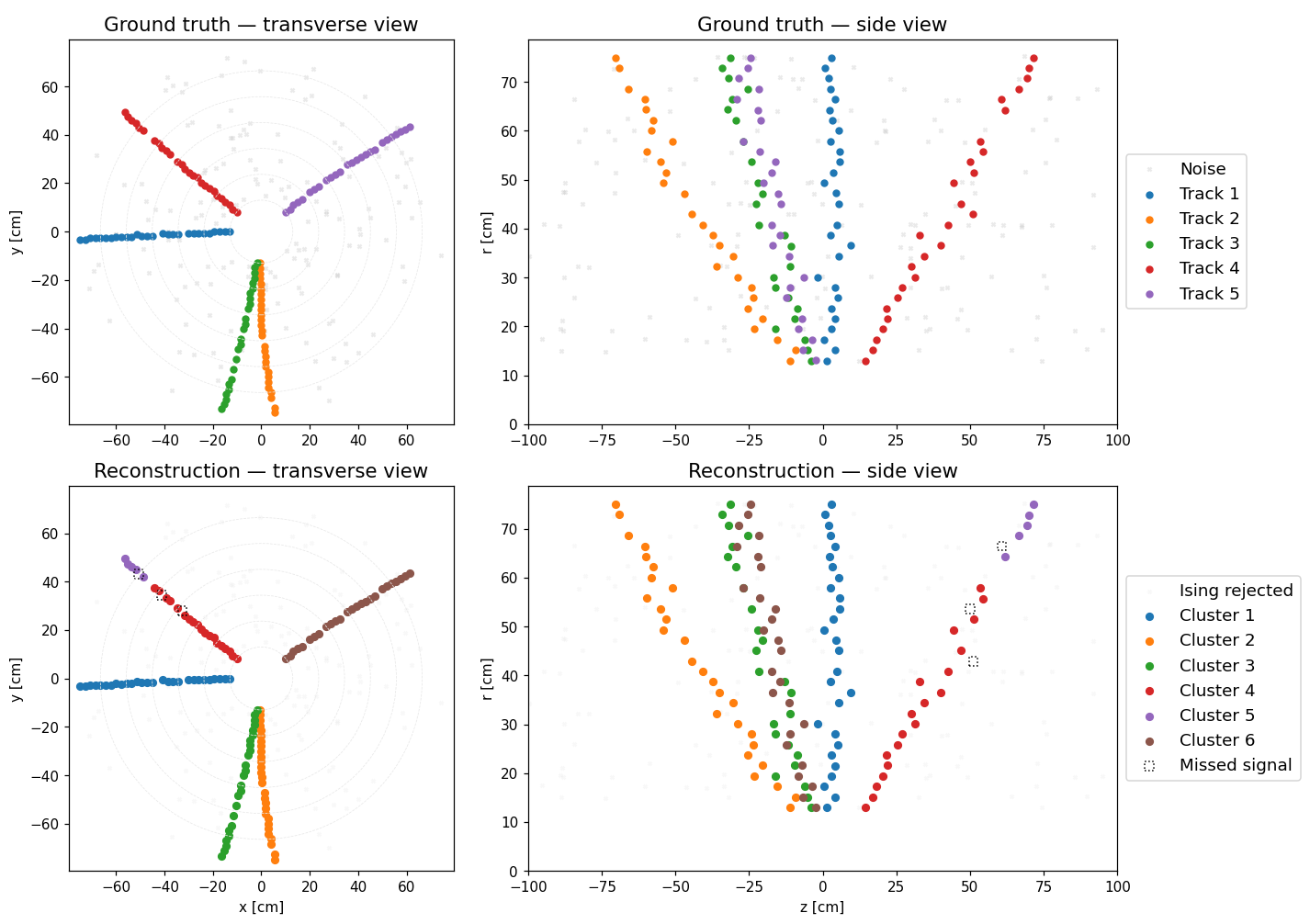}
    \caption{Illustration of applying the combination of Ising filter and Peterson network for tracks extraction.}
    \label{spd_tracking}
\end{figure}

Figure \ref{spd_tracking} illustrates the results of consecutive application of Ising filter and SPD-optimized Peterson tracking. The tracks are extracted by identifying the connected components of the tracks graph. 
 
Performance is quantified using the TrackML challenge score~\cite{Amrouche2019}, which measures the fraction of hits correctly assigned to reconstructed tracks. \chg{For this benchmark, we used single event read-out, with mean of 5 tracks per event.} The standard Peterson formulation, applied to unfiltered hits, achieves a score of \chg{$0.5$}. Applying the same standard coupling to Ising-filtered hits improves the score to \chg{$0.9$}, reflecting the reduction in fake doublets from the pre-filtered input. This demonstrates that the geometric physical understanding embedded in the Ising filter kernels can be directly extended to the track-finding stage. It is worth noting that applying the Ising filter prior to Peterson-inspired track finding \chg{speeds up the total execution time by a factor of about $2.5$}.
 
We observed that track extraction might fails in two cases. First, when some of the hits between track segments are missing, the track becomes fragmented into two parts. Second, overlapping tracks might get merged into a single one. These issues can be resolved by track post-processing on more realistic Monte Carlo data.

\section{Discussion}
\label{conclusion}

The results presented in this paper rest on a simple physical observation: signal hits produced by a single particle interaction are correlated by the laws of propagation — Cherenkov timing, helical geometry, or plain space-time locality — while noise hits are not. The Ising filter makes this observation quantitative by casting hit-level classification as energy minimization on a graph whose edge weights encode whichever propagation laws are relevant to the experiment at hand. The framework makes no assumption about global event topology, requires no preliminary track hypothesis, and converges in a handful of iterations.
 
Two experiments spanning very different physical regimes validate this picture. In Baikal-GVD, encoding Cherenkov propagation timing into the hit couplings achieves noise suppression performance comparable to the dedicated scan-fit algorithm. For the toy model of SPD straw tracker, where no timing information is available, purely geometric kernels yield $F_1 = 0.96$ for the signal hits selection. Furthermore, lifting the same constraints to the doublet level in the Peterson–Hopfield network raises the TrackML track-finding score from \chg{$0.5$ to $0.9$}. \chg{This demonstrates that using Ising filter is applicable in realistic and demanding cases.}
 
\chg{From a computational standpoint, the Ising filter is highly efficient, as all stages of the algorithm reduce to vectorized array operations. The coupling matrix $J_{ij}$ is constructed via pairwise computations on $N \times N$ arrays: all pairwise distances, time differences, and kernel evaluations are performed simultaneously, eliminating the need for an explicit loop over $O(N^2)$ hit pairs. When edges are restricted to $k$ nearest neighbors or a spatial cutoff, the coupling matrix becomes sparse, and all operations scale as $O(N \cdot k)$ per iteration. With the synchronous spin update rule, which yields the same performance metrics as the greedy update rule, each subsequent iteration reduces to a single matrix-vector multiplication $J \cdot \mathbf{s}$, providing full vectorization of the algorithm.}

\chg{Compared with neural-network classifiers trained for the same
tasks~\cite{kharuk2023rejecting, nikolskaiia2024point}, the Ising filter achieves
comparable precision and recall at a fraction of the computational cost. It also converges in
a few iterations with a handful of parameters instead of a deep forward pass with many
learned weights, and so remains practical on CPU-only systems. Those parameters are
also physically interpretable: each kernel encodes a specific propagation law, so they
can be constrained directly from measured detector properties rather than learned from
simulation. The filter is therefore less exposed to domain shift --- the degradation
a network trained on Monte Carlo suffers on real data, where effects such as
photosensor aging, seasonal background variations, and calibration drifts are not fully
simulated. This makes Ising filter also easier to maintain over the lifetime of an
experiment.}

One of the method's practical features is its portability. Adapting the Ising filter to a new experiment requires only a model of the dominant signal propagation process (for constructing the kernels $J_{ij}$) and realistic Monte Carlo sample (for hyperparameters optimization). In the domain of large-volume neutrino telescopes, the Cherenkov-propagation kernels developed here for Baikal-GVD apply directly to IceCube~\cite{IceCube2017}, KM3NeT~\cite{KM3NeT2016}, and the planned Baikal-GVD extension Baikal-HUNT~\chg{\cite{chen2026hunt}}. For collider tracking detectors, the helical-geometry kernels of Section~\ref{spd} are equally applicable to the inner tracker of any solenoidal experiment — candidate systems include the straw-tube tracker of NA62~\cite{NA62_2017} and other subdetectors of the NICA programme.  

\section*{Code availability}

The code for the Ising filter and Peterson network track finding is available at \small{\url{https://github.com/ml-inr/Ising-filter-and-tracking}}.

\section*{Acknowledgments}

This work was supported by the Russian Science Foundation under grant no. 24-72-10056.

The author thanks the Baikal-GVD collaboration for providing Monte Carlo samples and for useful discussions.

\bibliography{references}

\end{document}